\batchmode
\documentstyle[12pt]{article}

\newcommand{\bye}{\end{document}}

\newcommand{\sla}{\:\!\!\!\!\!/}
\newcommand{\be}{\begin{equation}}
\newcommand{\ee}{\end{equation}}
\newcommand{\bc}{\begin{center}}
\newcommand{\ec}{\end{center}}
\newcommand{\bes}{\begin{eqnarray}}
\newcommand{\ees}{\end{eqnarray}}
\newcommand{\nn}{\nonumber}
\newcommand{\nnn}{\nonumber\\}

\begin{document}

\setcounter{page}{0}
\pagestyle{empty}
\normalsize{
\hspace*{13.0cm}                                  CERN-TH.6869/93\\
\hspace*{13.0cm}                                  WU-B 93-06\\
}
\renewcommand{\thefootnote}{\fnsymbol{footnote}}
\Large
\bc
         {\bf  ON EXCLUSIVE REACTIONS IN THE TIME-LIKE REGION
\footnote[2]{Supported in part by the Bundesministerium
f\"{u}r Forschung und Technologie, FRG, under contract number 06 Wu 765
\newline
1) Theory Division, CERN, CH-1211 Geneva 23, Switzerland
\newline
2) Fachbereich Physik, Universit\"{a}t Wuppertal, Gau\ss strasse 20,
Postfach 10 01 27,\newline D-5600 Wuppertal 1, Germany
\newline
3) Institut f\"ur Theoretische Physik, Universit\"at Graz,
Universit\"atsplatz 5, A-8010 Graz, Austria
\newline
\newline
\newline
\newline
\normalsize
CERN-TH.6869/93
\newline
WU-B 93-06
\newline
April 1993
}}
\\
\vspace{1cm}

\ec
\bc
P. Kroll$^{1,2)}$, Th. Pilsner$^{2)}$, M. Sch\"urmann$^{2)}$ and W.
Schweiger$^{3)}$
\ec
\normalsize
\bc

                    {\bf Abstract}
\ec
The electromagnetic form factors of the proton in the time-like region and
two-photon annihilations into proton-antiproton are investigated. To
calculate these processes at moderately large $s$ we use a variant of the
Brodsky-Lepage hard-scattering formalism
where diquarks are considered as quasi-elementary constituents of baryons.
The proton wave function and the parameters controlling the diquark
contributions are determined from fits to space-like data.
We also comment on the decay $\eta_c \to p\bar{p}$.

\newpage
\normalsize
\pagestyle{plain}
\renewcommand{\thefootnote}{\arabic{footnote}}
There is general agreement that perturbative QCD in the framework of the
hard-scattering picture \cite{bro:80} can successfully make predictions
for exclusive reactions at asymptotically large momentum transfer $Q$.
However, the applicability of this scheme at experimentally accessible
momentum transfers was questioned \cite{isg:89,rad:91}. It was asserted
that in the few $\mbox{GeV}$ region there are strong contributions
from the end-point regions where one of the exchanged gluons becomes soft.
This
statement was recently disproved by Sterman and collaborators
\cite{bot:89,ste:92,li:92}. In the end-point regions the transverse
momenta of the quarks cannot be neglected although this is done customarily.
If one retains the transverse momenta, the gluon
virtualities are increased on the average and as a consequence the
contributions from
the dangerous end-point regions are diminished.
Additional suppression of the contributions from the end-point regions
is obtained from Sudakov form factors. It was shown for instance
that perturbative QCD
can readily be applied to the form factor of the pion for $Q^2 \geq
4\,\mbox{GeV}^2$
\cite{ste:92} and to the magnetic form factor of the proton
for $Q^2 \geq 12\, \mbox{GeV}^2$
\cite{li:92} in sharp contrast to the statements made in \cite{isg:89,rad:91}.
The numerical values for the form factors obtained in \cite{ste:92,li:92}
are almost the same as those found in the $\mbox{na}\ddot{\mbox{\i}}\mbox{ve}$
calculations when transverse
momenta and Sudakov suppressions are ignored. Thus the applicability
of the hard-scattering scheme seems to be justified in the few $\mbox{GeV}$
region.\\
In the light of this theoretical progress the new data on the
magnetic form factor $G_M^p$ of the proton in the time-like region can be
considered as a challenge. The data of the Fermilab
E760 collaboration \cite{arm:92}, shown in Fig. 1 together with the data
of~\cite{ff:91}, are about a factor of 3 larger than
the space-like data~\cite{arn:86} (at $Q^2=-s$). It seems very hard for
perturbative
QCD to reconcile the space- and time-like values of the form factors
around $10\,\mbox{GeV}^2$ in a theoretically self-consistent
way\footnote{Asymptotically,
perturbative QCD predicts the same values
for the form factors in the time- and space-like regions up to a
correction of order $\alpha_s$~\cite{Mag:90}.}.
Rather one may suspect that higher twist effects are responsible for
the difference. This conjecture is supported by many other observations
made in exclusive reactions at moderately large momentum transfer
(e.g. the difference between the magnetic and electric nucleon
form factors or polarizations).\\
There is another piece of data lacking theoretical explanation.
{}From CLEO \cite{cle:92} we have now at our disposal
precise data on two-photon annihilations into proton-antiproton.
The integrated cross-section ($|\cos\theta| \leq 0.6$) is well
measured up to $s=10.6\,\mbox{GeV}^2$ allowing for the first time a serious
comparison with theoretical predictions. It becomes evident that
the QCD predictions for that time-like process \cite{far:85}
are considerably below the data (about a factor of 8).
According to Hyer \cite{hye:92}
who has reconsidered the process $\gamma\gamma\to p \bar{p}$ recently,
an $s$-value of $10 \,\mbox{GeV}^2$
seems too small for making reliable predictions from QCD within the framework
of
the pure hard-scattering scheme.\\
The aim of the present paper is the explanation of these two sets of
new data. We will also comment on a third
time-like process, namely the decay $\eta_c \to p \bar {p}$. The
rather large decay width of about $10\, \mbox{keV}$ found by several
experimental groups \cite{eta:92,pdg:92} has not yet been understood.
To calculate the processes in question we use a variant of the
Brodsky-Lepage picture \cite{bro:80} for exclusive reactions in
which a baryon is assumed to consist of a quark and a diquark.
The diquark, being a cluster of two valence quarks
and a certain amount of glue and sea quark pairs, is regarded
as a quasi-elementary constituent, which partly survives medium
hard collisions. The composite nature of the diquark is taken
into account by diquark form factors (phenomenological $n$-point
vertex functions), which are parametrized in such a way that the pure quark
picture of Brodsky and Lepage emerges asymptotically.\\
The quark-diquark model of baryons has turned out to work rather
well for exclusive reactions in the space-like region
\cite{kro:90,kro:91}. It is particularly well suited for moderately
large momentum transfer since the diquark picture models non-perturbative
effects, in fact correlations in the baryon wave functions,
which are known to play an important role in that
kinematical region. With a common set of parameters specifying
the diquarks and process independent wave functions (or
distribution amplitudes, DAs) for the involved hadrons a good description
of a large number of exclusive reactions in the space-like region
has been accomplished by now. Among these reactions are the
electromagnetic form factors of baryons, Compton scattering off
protons and photoproduction of mesons. Diquarks have also been used
in many other soft and hard processes. For a review of recent
applications of the diquark picture see \cite{kro:93}.\\
In the hard-scattering model a form factor or a scattering amplitude
is expressed by a convolution of DAs with hard-scattering amplitudes
calculated in collinear approximation within perturbative QCD.
In a collinear situation in which intrinsic transverse momenta
are neglected and all constituents of a hadron have momenta
parallel to each other and parallel to the momentum of the parent
hadron, one can write the valence Fock state of the proton in a covariant
fashion \cite{geo:91} (omitting colour indices for convenience)
\be
\label{a1}
|p,\lambda\rangle  = f_S \Phi_S(x_1)B_S u(p,\lambda)
             + f_V \Phi_V(x_1) B_V
              (\gamma^{\alpha}+p^{\alpha}/m)\gamma_5 u(p,\lambda)/\sqrt{3}
\ee
$u$ is the spinor of the proton, $p$ and $m$ its momentum and mass,
respectively. The two terms in (\ref{a1}) represent configurations
consisting of a quark and either a spin-isospin zero $(S)$ or a
spin-isospin one $(V)$ diquark, respectively. The couplings of the diquarks
with the quarks in an isospin 1/2 baryon lead to the flavour functions
\be
\label{a2}
B_S=u\, S_{[u,d]}\hspace{2cm} B_V= [ u V_{\{u,d\}} -\sqrt{2} d\,
V_{\{u,u\}}]/\sqrt{3}\, .
\ee
The DA $\Phi_{S(V)}(x_1)$, where $x_1$ is the momentum fraction carried
by the quark, represents a light-cone wave function integrated over
transverse momentum. The constant $f_{S(V)}$ is the configuration space
wave function at the origin. \\
Representative Feynman diagrams contributing to the hard-scattering
amplitudes for the processes of interest in this article, are displayed in
Fig. 2. The gluon-diquark vertices are defined
by \cite{kro:90}
\bes
\label{a}
    SgS &: & +i g_s t^a (p_1 + p_2)_{\mu}  \nnn
    VgV &: & -i g_s t^a [ g_{\alpha\beta}(p_1+p_2)_{\mu}
                        - g_{\mu\alpha}\left[(1+\kappa_V)p_1-\kappa_V\,
p_2\right]_{\beta}
                        - g_{\mu\beta}\left[(1+\kappa_V)p_2-\kappa_V\,
p_1\right]_{\alpha}]\, ,
\ees
where $g_s=\sqrt{4\pi\alpha_s}$ is the QCD coupling constant.
$\kappa_V$ is the anomalous magnetic moment of the vector diquark and
$t^a(=\lambda^a/2)$ the Gell-Mann colour matrix. The rest of the notation
should be obvious. Gauge invariance also requires contact terms
($\gamma SgS$ and $\gamma VgV$). The form of these couplings is standard
and we refrain from repeating them here. \\
In applications of the diquark model, Feynman diagrams are calculated
with these rules for point-like particles. However, in order to take
into account the composite nature of the diquarks phenomenological
vertex functions have to be introduced. Advice for the parametrizations
of the 3-point functions, ordinary diquark form factors, is obtained
from the requirement that asymptotically the diquark model evolves
into the pure quark model of Brodsky and Lepage. In view of this
the diquark form factors in the space-like region are parametrized as
\bes
\label{a4}
  F_{S}(Q^2)&=& F^{(3)}_{S}(Q^2)=1/(1+Q^2/Q^2_{S}) \nnn
  F_{V}(Q^2)&=& F^{(3)}_{V}(Q^2)=1/(1+Q^2/Q^2_{V})^2 \, .
\ees
In accordance with the required asymptotic behaviour the $n$-point
functions ($n\geq 4$) are parametrized as
\be
\label{a5}
F^{(n)}_S(Q^2)=a_S F_S(Q^2) \qquad
            F^{(n)}_V(Q^2)=a_V F_V(Q^2)/(1+Q^2/Q^2_V)^{(n-3)}\, .
\ee
The constants $a_{S,V}$ are strength parameters, which take into account
absorption due to diquark excitation and break-up.\\
Putting all this together we find for the magnetic form factors of the
proton in the time-like region:
\bes
\label{a6}
&&G_M^p=\frac{4\pi}{3}C_F\int_0^1 dx_1 dy_1
\Bigl[-2\frac{f_S^2}{s}\Phi_S(y_1) T_S(x_1,y_1,s) \Phi_S(x_1)
   +\frac{f_V^2}{m_V^2}\Phi_V(y_1) T_V(x_1,y_1,s) \Phi_V(x_1)\Bigr] \nnn
&& T_S(x_1,y_1,s)= 2\frac{\alpha_s(\tilde{s}_{22})}{x_2 y_2}
   F_S(\tilde{s}_{22})
                 +\frac{\alpha_s(\tilde{s}_{11})}{x_1y_1}
   F_S^{(4)}(\tilde{s}_{11}+\tilde{s}_{22}) \nnn
&& T_V(x_1,y_1,s)=\frac{\alpha_s(\tilde{s}_{11})}{x_1y_1}
   F_V^{(4)}(\tilde{s}_{11}+\tilde{s}_{22})
\Bigl(\kappa_V^2(1+2\,x_1+2\,y_1
+x_1y_1)+\frac{1}{2}\kappa_V(1-\kappa_V)(x_2+y_2)\Bigr.\nnn
&&\hspace{2cm}\;\Bigl.-\frac{3}{2}x_2 y_2(1-\kappa_V^2)\Bigr)
\ees
where $\tilde{s}_{ij}=x_i y_j s$ and $x_2(y_2)=1-x_1(y_1)$;
$C_F =4/3$ is the colour factor and
$m_V$ the mass of the vector diquark.
It appears in the vector diquark propagator.
In contrast to the pure quark model we can also calculate
the Pauli form factor:
\bes
\label{c2}
F_2^p&=&\frac{8\pi C_F f_V^2 }{3 \kappa_p m_V^2}
      \int_0^1 dx_1 dy_1 \Phi_V(y_1) T_V(x_1,y_1,s)\Phi_V(x_1)\nnn
T_V(x_1,y_1,s)&=&\frac{\alpha_s(\tilde{s}_{11})}{x_1y_1}
      F_V^{(4)}(\tilde{s}_{11}+\tilde{s}_{22})
      \kappa_V\left((\kappa_V+\frac{1}{2})(x_1+y_1)+\kappa_V-1\right)
\ees
and hence the electric form factor
\be
\label{aa7}
       G_E^p=G_M^p -\kappa_p (1-s/(4m^2)) F_2^p \, .
\ee
The process $\gamma\gamma\to p \bar{p}$ is described by six independent
helicity amplitudes $M_i$~\cite{kro:91} in terms of which
the differential cross-section is expressed by:
\be
\label{a9}
\frac{d\sigma}{d\Omega}=\frac{\sqrt{1-4m^2/s}}{64\pi^2 s}\,
\frac{1}{2}\left[|M_1|^2+|M_2|^2+2\,|M_3|^2+2\,
|M_4|^2+|M_5|^2+|M_6|^2\right]\, .
\ee
Each amplitude is a sum of the two contributions from scalar $(M_i^S)$
and vector $(M_i^V)$ diquarks.
These contributions can be written as $(D=S,V)$
\be
\label{n1}
  M_i^{D}=\frac{64\pi^2 C_F \alpha}{9\sqrt{tu}} f_{D}^2
           \int_0^1 dx_1 dy_1 \frac{\Phi_D(y_1)\Phi_D(x_1)}{x_1 y_1}
\hat{M}_i^{D}\, .
\ee
For scalar diquarks, only the two helicity non-flip amplitudes, $\hat{M}_1^S $
and $\hat{M}_5^S$, are non zero. These amplitudes read:
\bes
\label{a7}
& &\hat{M}_1^S(s,t,u)= \hat{M}_5^S(s,u,t)=
4\alpha_s(\tilde{s}_{22}) F_S(\tilde{s}_{22})\frac{x_1y_1t +x_2y_2 u}{x_2 y_2
s}\\
&&-2\alpha_s((\tilde{s}_{12}+\tilde{s}_{21})/2)
           F_S^{(4)}((\tilde{s}_{12}+\tilde{s}_{21})/2+\tilde{s}_{22})
          \frac{2x_1y_1x_2y_2s^2 +(x_1-y_1)^2tu+x_1y_1(x_2+y_2)st}
            {(x_2y_1u+x_1y_2t)(x_2y_1t+x_1y_2u)}\nnn
& & -\alpha_s(\tilde{s}_{11}) F_S^{(5)}(\tilde{s}_{11}+\tilde{s}_{22})\, .  \nn
\ees
The vector diquark contribution to the helicity amplitudes form very
lengthy expressions and we refrain form quoting them in full
(see however,~\cite{kro:91}). Rather we restrict ourselves to
the contributions, by far dominant from the 3-point function, which read:
\be
\label{a8}
\hat{M}_1^V(s,t,u)= \hat{M}_5^V(s,u,t)=\frac{\kappa_V}{3 x_2 y_2 m_p^2}
     \alpha_s(\tilde{s}_{22}) F_V(\tilde{s}_{22})(x_1 y_1 t +x_2 y_2 u)\, .
\ee
The other four helicity amplitudes contribute less than $10\%$ to the
differential cross-section.
The space-like versions of the expressions (\ref{a6}-\ref{a8})
have been published by us \cite{kro:90,kro:91}. The
time-like and space-like expressions are related to each other by
$t (=-Q^2) \leftrightarrow s$ crossing. For details of the
calculations and a discussion of the various assumptions
underlying these calculations, we refer to \cite{kro:90,kro:91}.
We note that there are modifications of minor importance in the
above expressions.
These
modifications arise from the use of the covariant spin wave
functions (\ref{a1}). For the vector diquark contributions this leads
to some correction terms related to helicity flips of the quark.
Such terms have been neglected in our previous work. The use of
covariant spin wave functions has many technical advantages.
For instance the calculation of a large set of so-called
elementary amplitudes is avoided, one immediately projects onto
hadronic states. Another advantage is that only hadronic quantities
(spinors, polarization vectors, and so on) appear.
The covariant spin wave functions
are particularly well suited for algebraic computer programs, the
use of which is unavoidable in extensive calculations such as the
one required for the process $\gamma \gamma\to p \bar{p}$. These
modifications due to the use of the covariant
spin wave functions lead to small changes of the parameters.
We have therefore repeated our fits for the electromagnetic form factors of the
nucleon (see Fig. 1) and found similarly good fits to the
data as in our previous studies \cite{kro:91,kro:93} with the following proton
DAs
\be
\label{a10}
\begin{array}{l}
\Phi_S(x_1)=N_S x_1 x_2^3\exp{\left[-b^2 (m^2_q/x_1+m^2_S/x_2)\right]}\\
\Phi_V(x_1)=N_V x_1 x_2^3(1+5.8 x_1-12.5 x_1^2)
\exp{\left[-b^2 (m^2_q/x_1+m^2_V/x_2)\right]}
\end{array}
\ee
and the set of parameters
\be
\label{c1}
\begin{array}{cccc}
 f_S= 73.85\,\mbox{MeV}, & Q_S^2=3.22 \,\mbox{GeV}^2, & a_S=0.15, &  \\
 f_V=127.7\,\mbox{MeV}, & Q^2_V=1.50 \,\mbox{GeV}^2, & a_V=0.05,  &
\kappa_V=1.39\,;
\end{array}
\ee
$\alpha_s$ is evaluated with $\Lambda_{QCD}=200\,\mbox{MeV}$. The
parameters $Q_S$ and $Q_V$, controlling the size of the diquarks,
are in agreement with the higher twist effects observed in deep
inelastic lepton-hadron scattering \cite{dis:91} if these effects
are modeled as lepton-diquark elastic scattering. The DAs are
a kind of harmonic oscillator wave function transformed to the light-cone.
The masses in the exponentials are constituent masses since they enter
through a rest frame wave function. For the quarks we take $330 \,\mbox{MeV}$
whereas for the diquarks $580 \,\mbox{MeV}$ is used. The oscillator parameter
b is taken to be $0.248\,\mbox{GeV}^{-1}$. The constant $N_{S(V)}$ is fixed by
the
usual auxiliary requirement
$\int dx_1 \Phi_{S(V)}(x_1)=1$ ($N_S=25.97; N_V=22.29$). The more
complicated form of the DA $\Phi_V$ causes a smaller mean value of $x_1$
than obtained for the DA $\Phi_S$.
The exponentials in the DAs
provide strong suppressions in the end-point regions. Consequently a treatment
of these regions in the manner proposed by \cite{bot:89,ste:92,li:92}
does not change the results much\footnote{We thank R. Jakob for informing us on
this result.}.
In view of this and since there is always one gluon less in the diquark model
than in the pure quark model where the proton consists of three
valence constituents instead of two,
the diquark model can be applied for
momentum transfers $\geq 4 \mbox{GeV}^2$.\\
Before confronting the model with experiment a remark concerning the
diquark form factors is in order. The relations (4) represent an
effective parametrization valid at large space-like $Q^2$.
It is not possible to continue these parametrizations in a unique
way to the time-like region since we do not know the exact dynamics
of the diquark system. We define a continuation to the time-like
region as follows: $Q^2$ is replaced by $-s$ in eq. (4). The
correct asymptotic behaviour is so guaranteed. This simple
replacement however leads to poles in the diquark form factors
which have no real physical meaning. In order to avoid these unphysical
poles we keep the diquark form factors constant once they have
reached a certain value, say $c_0$. A similar recipe has been used
by Ekelin et al.~\cite{eke:85} in a study of $e^+ e^-$ annihilation
into hadrons. Our
results are not very sensitive to the value of $c_0$ since the cut-off of
the diquark form factors is only effective in the end-point regions. We
expect the role of this parameter to diminish further if Sudakov
form factors are taken into account.\\
With these remarks our model is fully specified and we can compare
our predictions with data. We find good agreement with the data
for $c_0=1.3\, $. In Fig. 1 we present our predictions for the magnetic form
factor
of the proton. Good agreement is obtained with the data of the E760
collaboration
\cite{arm:92}. $G_M$ does not yet behave as $1/s^2$. It falls down somewhat
faster,
approaching slowly the value of the space-like form factor. Typical
vector meson dominance models cannot explain the large difference
between the values of the time-like and space-like form factors in the
$6-15\,\mbox{GeV}$ region. The electric form factor $G_E$ behaves
in a way similar to $G_M$,
but has a value about $15\%$ larger than that of $G_M$ for $s$ between 6 and
$15\,\mbox{GeV}$.
Armstrong et al. \cite{arm:92} assumed $|G_E|=|G_M|$ in the analysis of their
$e^+e^-\to\bar{p}p$ data. Our model provides justification for this
assumption. We note that there is no prediction  for $G_E$ from the pure
quark model.\\
In Fig. 3 we display the integrated cross-section
($|\cos\theta|\leq 0.6$) for two-photon annihilations into
$p\bar{p}$ as a function of $\sqrt{s}$. For large values of
$\sqrt{s}$ ($\geq 2.5\,\mbox{GeV}$) we find good agreement with the CLEO data
\cite{cle:92}, whereas the pure quark model fails by about a factor
of 8~\cite{far:85}. The differential cross-section is found to
increase smoothly from $90^0$ downwards smaller angles.
For comparison we quote
the values of the cross-section integrated over several angular
ranges (at $\sqrt{s}=3 \,\mbox{GeV}$):
\be
\label{a11}
\sigma (\gamma\gamma\to p\bar{p})=\left\{\begin{array}{rcl}
127\,\mbox{pb} &\mbox{for} & |\cos\theta|\leq 0.6 \\
 73\,\mbox{pb} &\mbox{for} & |\cos\theta|\leq 0.4 \\
 34\,\mbox{pb} &\mbox{for} & |\cos\theta|\leq 0.2 \end{array}\right.
\ee
The last process we are interested in is the decay of the $\eta_c$ into
$p\bar{p}$. The amplitudes for that process have the covariant
structure
\be
\label{b1}
   M(\eta_c\to p\bar{p})= B_{\eta_c}(M^2)\bar{u}(p_1)\gamma_5 v(p_2)\, ,
\ee
where $M$ denotes the meson mass. The decay width for that channel
is expressed in terms of the invariant function $B_{\eta_c}$ by
\be
\label{b2}
      \Gamma(\eta_c\to p\bar{p})=
                   \frac{1}{8\pi}\sqrt{M^2-4m^2} |B_{\eta_c}|^2\, .
\ee
The invariant function $B_{\eta_c}$ is to be calculated from diagram
2c and the one with the two gluons interchanged. We obtain:
\bes
\label{b7}
&&B_{\eta_c}=i\frac{128\pi^2}{3\sqrt{2}}C_F\frac{M^2}{m}(M^2-4m^2)
          f_V^2 f_{\eta_c} \int_0^1 dx_1 dy_1 dz_1 \Phi_V(y_1)\Phi_V(x_1)
         \Phi_{\eta_c}(z_1)  \nnn
        & & \times\,\frac{\alpha_s(g_1^2)\alpha_s(g_2^2)}{g_1^2 g_2^2}
F_V(g_2^2)
         \frac{(x_1-y_1)^2}{(z_1-x_1)(z_1-y_1)M^2+(x_1-y_1)^2 m^2-m_c^2} \, ,
\ees
where $g_i^2=(x_i-y_i)^2 m^2+x_i y_i M^2$.
The scalar diquark contribution is zero. The pure
quark model of Brodsky and Lepage also provides a zero
$\eta_c\rightarrow p\bar{p}$ width. Eventual propagator poles in
eq. (\ref{b7}) are treated with the usual $i\varepsilon$-method.
The colour factor $C_F$ is $2\sqrt{3}/9$ and $m_c$ is the $c$-quark mass.\\
In deriving eq. (\ref{b7}) we have written the $\eta_c$ state in a
covariant fashion, as we did for the proton
\be
\label{b4}
          f_{\eta_c} \Phi_{\eta_c}(z_1)(P\sla +M)\gamma_5/\sqrt{2}
\ee
where $P$ denotes the meson momentum. For the $\eta_c$
wave function we use the Bauer-Stech-Wirbel ansatz~\cite{wir:85}
\be
\label{b9}
       \Psi(z_1,k_{\perp})= \tilde N \Phi_{\eta_c}(z_1)
                            \exp[-b_{\eta_c}^2 k_{\perp}^2]
\ee
where the DA is given by
\be
\label{b10}
       \Phi_{\eta_c}(z_1)=N_{\eta_c} z_1z_2
\exp{\left[-b_{\eta_c}^2 M^2 (z_1-1/2)^2\right]}\, .
\ee
The DA satisfies the usual
condition $\int dz_1 \Phi_{\eta_c}(z_1)=1$ ($N_{\eta_c}=9.59$).
Its $z$-dependence is characteristic of a $c\bar{c}$ meson:
It is symmetric in $z_1$ and $z_2$ and strongly peaked at $z_1=1/2$.
Since for heavy hadrons the valence Fock state is expected to
be dominant, the wave function (\ref{b9}) is normalized to
1 leaving as the only free parameter the oscillator parameter
$b_{\eta_c}$. Using the information on the radius obtained from
non-relativistic potential studies of the charmonium family, we
can fix the oscillator parameter. We take
$\langle r^2\rangle ^{1/2}_{\eta_c}=\langle r^2\rangle ^{1/2}_{\Psi}
=0.42\,\mbox{fm}$ and obtain $b_{\eta_c}=1.21\,\mbox{GeV}^{-1}$.
The wave function (\ref{b10}) is thus entirely fixed and hence the constant
$f_{\eta_c}$,
for which we get the reasonable value of $70.7\,\mbox{MeV}$.
As a check of the $\eta_c$ wave function we calculate
the width for the decay into two photons.
With the DA (\ref{b10}), the above value for $f_{\eta_c}$, and
the value of $1.35 \, \mbox{GeV}$for the current mass of the $c$-quark  we
obtain $\Gamma(\eta_c\to\gamma\gamma)=7.21\,\mbox{keV}$, which compares
favourably with the experimental value of $6.4{+2.4\atop-2.1}\,\mbox{keV}$
(PDG average \cite{pdg:92}).
Having convinced ourselves that the description of the $\eta_c$ is reasonable,
we insert everything into (\ref{b7}), perform the integrals and arrive at the
prediction
\be
\label{b12}
            \Gamma(\eta_c\to p\bar{p})=3.88\,\mbox{keV} \, .
\ee
In order to obtain that result we have taken into account admixtures
of light $q\bar{q}$ states in the $\eta_c$ according to  the estimate
grown by Fritzsch and Jackson~\cite{Fri:77}. These admixtures lead to an
increase of the width by about $20\%$.
We, however, present this result with some reservations since it is subject
to rather large theoretical uncertainties. Thus, for instance, it
strongly depends on the shape of the proton DA\footnote{The factor
$(x_1-y_1)^2$
appearing in (\ref{b7}) implies that the invariant function $B_{\eta}$ is
related
to the dispersion of the proton's DA,
$\langle x_1^2\rangle -\langle x_1\rangle ^2$.}.
Radiative corrections are probably very large~\cite{bar:76}. Eventual
admixtures
of pure glue states in the $\eta_c$ are
another source of uncertainties.
The experimental values of $\Gamma (\eta_c\to p \bar p)$ are
\cite{eta:92}: $10.3\pm 6.3\,\mbox{keV}$ (DM2), $11.3\pm 7.3\,\mbox{keV}$
(MARK3).
Despite the theoretical uncertainties,
the diquark model provides the right magnitude for the decay width.
In this respect we disagree
with the authors of \cite{ans:91}: there, the
difference between space-like and time-like diquark form factors
has been overlooked. \\
Along the same lines we can also calculate the width for the
processes $\eta_c'(3594)\rightarrow p\bar{p}$ and
$\eta_b(9600)\rightarrow p\bar{p}$. We find
$\Gamma (\eta_c'\to p \bar p)=0.13\,\mbox{keV}\;(m_c=1.35\,\mbox{GeV};
\langle r^2\rangle ^{1/2}_{\eta_c'}=0.85\,\mbox{fm})$ and
$\Gamma (\eta_b\to p \bar p)=8.5\cdot 10^{-3}\,\mbox{eV}\;(m_b=5\,\mbox{GeV};
\langle r^2\rangle ^{1/2}_{\eta_b}=0.23\,\mbox{fm})$.\\
There are a few other decay processes, which in principle can also
be calculated within the diquark model. The process
$\Psi\to p\bar{p}\gamma$ has been studied by Carimalo and Ong
\cite{car:91}. These authors obtain a decay rate that is
too small with respect to the data.
We expect that with the time-like version of the diquark
form factors one will find a larger decay rate.
Of considerable theoretical
interest are also the decays of $P$-wave quarkonia into $p \bar p$.
A hybrid approach has been used in previous investigations:
The $P$-wave states are treated non-relativistically whereas the light-cone
formalism is employed for the final state baryons. The quality of this approach
is unknown.
Moreover, as pointed out recently by Bodwin et al. \cite{bod:92},
strong contributions from higher Fock states are to be expected.
Therefore we believe that the $P$-wave charmonium decays
require a more detailed investigation
than is possible in a short article like this. \\
We summarize: The quark-diquark model successfully applies to many exclusive
reactions in both the space-like and the time-like regions. With
the DAs of the proton and the diquark parameters as fixed by fits to the
space-like
data and with a plausible treatment of the diquark form factors in the
time-like region,
we can explain the data for the proton form factor in the time-like region and
those
for the $\gamma\gamma \to p \bar p$ cross-section. We even obtain a
reasonable value for the $\eta_c \to p \bar p$ decay width. Finally, we note
that
in the time-like region we trust the predictions of the diquark model for
$s \geq 6\, \mbox{GeV}^2$.\\ \\
Acknowledgement: We thank K. Seth for discussions of the E760 results.

\newpage
\\
{\large {\bf FIGURE CAPTIONS} }
\\
\newline
Fig. 1 The magnetic form factor of the proton in the time-like and
space-like (at $Q^2=-s$) regions. The solid lines represent the
predictions of the diquark model. The time-like data $(\circ,\Box)$
are taken from Ref.~\cite{arm:92,ff:91}, the space like-data $(\bullet)$
from Ref.~\cite{arn:86}.
\\
Fig. 2 Typical Feynman diagrams contributing to the electromagnetic
form factor (a), to the $\gamma \gamma$ annihilation into $p \bar{p}$ (b)
and to the decay $\eta_c\to p\bar{p}$ (c).
\\
Fig. 3 The integrated cross-section ($|\cos\theta|\leq 0.6$) for
$\gamma\gamma\to p\bar{p}$ as a function of $\sqrt{s}$. Data are taken from
\cite{cle:92}.
\\
\bye